\documentclass[preprint,12pt,authoryear]{elsarticle}

\usepackage{amssymb}
\usepackage{amsmath}
\usepackage{tabularx}
\usepackage{rotating}
\usepackage{url}
\usepackage{tabularx}
\usepackage{pdflscape}
\usepackage{ulem}
\usepackage{booktabs} 

 \usepackage{lineno}

\journal{Earth Science Reviews}

\begin{document}

\begin{frontmatter}



\title{A Unified Conceptual Framework for Gravitational Instabilities: From Latent System State to Failure}

\author[label1]{Faillettaz Jerome} 

\affiliation[label1]{organization={Cerema, Univ Gustave Eiffel, UMR Telluris},
            city={Bron},
            postcode={F-69674}, 
            country={France}}

\begin{abstract}
Gravitational instabilities are among the most widespread natural hazards and are expected to become increasingly significant under ongoing environmental change. Despite substantial advances in process understanding, monitoring, and modeling, predicting the timing of failure remains a fundamental challenge because instability emerges from complex interactions operating across multiple spatial and temporal scales. Existing approaches often focus on specific processes, forcing mechanisms, or observational signatures, resulting in a fragmented view of how systems evolve toward failure. This article propose a unified conceptual framework in which gravitational instabilities are interpreted as the progressive evolution of a latent system state controlled by damage accumulation, stress redistribution and external forcing to catastrophic failure. Within this perspective, failure emerges from the continuous interplay between internal system dynamics and external forcing, while predictability depends on our ability to infer the evolving state of the system from incomplete and indirect observations. The framework provides a common structure linking physical processes, observable manifestations, monitoring strategies, modeling approaches, and forecasting methods across a wide range of gravitational hazards. By integrating concepts from geomechanics, fracture mechanics, statistical physics, and data-driven sciences, the proposed framework shifts the focus from the search for universal precursors toward the reconstruction of evolving system states and their proximity to instability. It offers a unifying perspective for understanding failure processes and developing more robust approaches to hazard assessment and early warning.
\end{abstract}

%
%
%
%
%


\begin{keyword}


Gravitational instabilities \sep Landslides \sep  Rockfalls \sep  Glacier instabilities \sep  Failure prediction \sep  Damage mechanics; Early warning systems; Monitoring; State estimation; Critical transitions
\end{keyword}

\end{frontmatter}



\section{Introduction: The Prediction Problem}

Gravitational instabilities, including landslides, rockfalls, debris flows, glacier collapses, and slope failures in natural or engineered environments, represent one of the most widespread and destructive classes of natural hazards. These phenomena affect mountainous, volcanic, permafrost, and coastal regions worldwide, causing substantial human, economic, and environmental losses each year \citep{Crosta2008,Petley2012,IPCC2022,IPCC2022mountain}. Their occurrence is expected to increase in many regions under the combined influence of climate change, permafrost degradation, land-use modification, and expanding infrastructure development in unstable terrains \citep{Casagli&al2023}. Beyond their societal importance, gravitational instabilities provide a natural framework for investigating how heterogeneous geomaterials evolve from distributed deformation toward catastrophic failure across multiple spatial and temporal scales.

Despite major advances in geomechanics, remote sensing, monitoring technologies, and numerical modelling, predicting the timing of gravitational failure remains a fundamental scientific challenge. One major difficulty arises from the intrinsic heterogeneity of geomaterials, whose mechanical behaviour is controlled by complex interactions among fractures, pores, lithological variability, fluids, temperature, and evolving stress conditions \citep{Scholz2002,Alava&al2006}. Failure rarely results from the activation of a single discontinuity or isolated mechanism. Instead, instability emerges progressively through the collective evolution of interacting damage processes, stress redistribution, frictional weakening, and strain localization across multiple scales \citep{Ben-zion2008,Kun&al2014,Faillettaz&al2010}. This evolution is inherently nonlinear and may involve threshold effects, metastable states, and abrupt transitions between apparently stable and unstable configurations. Together, these processes define the evolving internal condition of the system, although this condition cannot be observed directly.

Current approaches remain largely fragmented across disciplinary perspectives. Continuum geomechanical models effectively describe stress evolution and deformation but often struggle to reproduce the emergence of discrete rupture and collective dynamics near failure. Discrete and fracture-based approaches explicitly represent crack propagation and localization processes but remain difficult to upscale to natural systems \citep{Jing2003,Stead&al2006}. Statistical physics models successfully reproduce scaling laws, avalanche dynamics, and critical transitions in heterogeneous media, yet generally rely on simplified geometries and boundary conditions \citep{Alava&al2006,Sornette2006,Pradhan&al2010}. Monitoring and remote-sensing techniques provide increasingly detailed observations of deformation, seismicity, hydrology, and thermal evolution, but these measurements remain indirect manifestations of the underlying mechanical evolution \citep{Casagli&al2023}. More recently, machine-learning and data-driven approaches have demonstrated remarkable capabilities for detecting patterns and forecasting instabilities from large datasets, although their physical interpretability often remains limited \citep{Reichstein2019}.

Similar observable signals—including accelerating deformation, evolving seismic activity, and hydrological or thermal changes—have been reported prior to failure across a wide range of gravitational systems. However, their interpretation depends strongly on system characteristics, forcing conditions, and observational scale, and no single observable or modelling approach provides a universal basis for prediction \citep{Intrieri2019}. The central difficulty is that monitoring systems only record indirect manifestations of ongoing processes rather than the internal mechanical evolution itself. Predictability therefore depends not only on identifying precursory signals but on inferring the evolving \emph{latent system state} of a heterogeneous geomaterial and its proximity to instability from incomplete, indirect, and uncertain observations.

Existing reviews have generally focused on specific hazards, monitoring technologies, numerical methods, or triggering mechanisms. Here, rather than reviewing these components independently, we propose an integrative conceptual framework that places the \emph{latent system state} at the centre of gravitational instability research. Within this perspective, the latent system state represents the internal mechanical condition of the geomaterial, integrating the cumulative effects of damage accumulation, stress redistribution, frictional evolution, localization, and external forcing. Although it cannot be observed directly, it governs the emergence of measurable signals and ultimately controls the transition toward failure.

The proposed framework links physical processes, observable manifestations, monitoring strategies, modelling approaches, and forecasting methods within a common conceptual structure. It integrates concepts from geomechanics, fracture mechanics, statistical physics, monitoring sciences, and data-driven approaches by explicitly relating internal system evolution to observable behavior and predictability. Rather than searching for universal precursors, this perspective shifts the emphasis toward reconstructing the evolving latent system state and estimating its proximity to instability. In doing so, it provides a unified basis for interpreting diverse gravitational hazards and for developing more physically grounded approaches to hazard assessment, forecasting, and early warning.

\section{Existing conceptual frameworks for failure in natural media}
\label{CCF}
Understanding the transition from stable deformation to catastrophic failure remains a central challenge in Earth system science. Over the past decades, several conceptual frameworks have been developed to explain failure processes in geomaterials and natural slopes. These frameworks differ fundamentally in their representation of material behavior, temporal evolution, and predictability, and they often emphasize distinct physical mechanisms. Rather than converging toward a single theory, the field has evolved into a set of partially overlapping paradigms, each supported by specific observational and experimental evidence. This section critically reviews the principal frameworks and evaluates their respective explanatory and predictive capabilities.

\subsection{Critical point and damage mechanics approaches}

The critical point framework interprets failure as the internal culmination of progressive damage accumulation within a heterogeneous medium \citep{Sornette2006}. Rooted in statistical physics and fracture mechanics, this approach describes failure as an emergent collective process in heterogeneous media \citep{Alava&al2006}.
As damage localizes and correlations grow, the system approaches a critical state characterized by scale-invariant behavior, including power-law acceleration of deformation or seismic activity.

Empirical support for this framework is found in numerous case studies reporting accelerating precursory signals prior to failure \citep{Voight1988,Sornette2006}. For example, time-to-failure models based on inverse velocity or power-law acceleration have been successfully applied to landslides, glacier break-off and rock failures, suggesting that macroscopic observables may reflect underlying critical dynamics \citep{SorAnd1998,Faillettaz&al2016b}. These behaviors are commonly interpreted as signatures of increasing correlations within the damaged medium.

However, the critical point paradigm remains debated. A major limitation is non-uniqueness: similar accelerating signals can arise from different physical processes, including purely kinematic effects or external forcing. Moreover, not all systems exhibit clear critical behavior; abrupt failures with limited precursors challenge the universality of this framework.
The critical-point approach therefore provides a powerful framework for interpreting emergent behavior, but its predictive value depends strongly on system conditions and the development of sufficient damage \citep{Ben-zion2008,Sornette2006}.

\subsection{Rate-and-state friction and frictional instability}

An alternative framework focuses on frictional processes governing slip along pre-existing or evolving shear zones . Rate-and-state friction laws describe how frictional resistance depends on slip velocity and the evolving state of contact surfaces \citep{Dieterich1979,Ruina1983}. Instability arises when velocity-weakening behavior leads to runaway acceleration, analogous to nucleation processes observed in earthquakes.

Applications to landslides and slow-moving slopes have demonstrated that rate-and-state friction can reproduce both stable creep and catastrophic failure within a unified formulation \citep{Helmstetter&al2004,Faillettaz&al2010}. Notably, this framework explains the coexistence of long periods of quasi-stable motion and sudden transitions to rapid failure, depending on system parameters such as stiffness, loading rate, and hydrological conditions. Observations of slow-moving landslides transitioning to rapid collapse are consistent with such frictional instabilities, where changes in pore pressure or effective stress modulate frictional properties \citep{Iverson2000}.

Despite its strong physical basis, the frictional framework often assumes the existence of a well-defined shear surface, which may not be present in highly fragmented or diffuse failure zones. Furthermore, the approach primarily captures localized instability, whereas many natural systems exhibit distributed damage prior to failure. Frictional models may thus underestimate the role of bulk material degradation and long-range interactions \citep{Ben-zion2008}.

\subsection{Threshold and hydromechanical triggering models}

A third class of models emphasizes the role of external forcing and threshold conditions. In these approaches, failure occurs when a controlling variable—such as pore pressure, rainfall intensity, or seismic loading—exceeds a critical threshold \citep{Iverson2000,Crosta2008}. Hydromechanical coupling is particularly important in slope stability, where transient infiltration can reduce effective stress and trigger failure \citep{Iverson2000,Wei&al2020}.

These models remain widely used because of their operational simplicity and empirical robustness \citep{Guzzetti&al2007}. Rainfall thresholds, for instance, have been successfully implemented in early warning systems across diverse climatic and geological settings \citep{Fan&al2015}. Similarly, hydro-mechanical models incorporating soil moisture dynamics and previous conditions provide physically grounded predictions of failure timing  \citep{Fan&al2015,Wei&al2020}.

However, threshold-based approaches are inherently limited in their ability to capture internal system evolution. They typically neglect progressive damage and assume that failure is externally triggered rather than internally driven. This can lead to false positives (threshold exceeded without failure) or false negatives (failure below threshold), particularly in systems where material degradation plays a dominant role. Consequently, threshold models are most effective when combined with information about internal state variables.

\subsection{Structural and kinematic controls}

Beyond constitutive behavior and external forcing, structural and geometrical factors exert a first-order control on failure processes. These include lithological heterogeneity, fracture networks, slope geometry, and the presence of pre-existing weaknesses. In many cases, such features determine the location, style, and scale of failure, independent of the specific triggering mechanism \citep{Stead2015,Lacroix&al2020}.

Kinematic analyses and limit equilibrium methods have long been used to assess slope stability based on these structural factors \citep{Hoek1981}. While these approaches provide valuable insights into potential failure modes, they are generally static and do not account for temporal evolution \citep{Lacroix&al2020}. As a result, they offer limited predictive capability regarding the timing of failure.

Recent advances attempt to integrate structural controls with dynamic processes, for example through stochastic representations of material heterogeneity or coupled mechanical-hydrological simulations \citep{Li&al2023}. These efforts highlight the importance of considering both initial conditions and evolving processes, bridging the gap between purely geometric and fully dynamic models \citep{Lacroix&al2020}.

\subsection{Synthesis: complementarities and limitations}
These frameworks are complementary rather than competing. Critical-point approaches emphasize collective dynamics and precursory organization, frictional models describe localized slip instability, threshold approaches focus on environmental forcing, and structural analyses constrain susceptibility and failure geometry. The main characteristics, strengths and limitations of these conceptual frameworks are summarized in Table \ref{tab:frameworks}. Their relative importance varies across systems and stages of destabilization. Predictability therefore depends on the interaction between damage evolution, frictional weakening, external forcing, and structural inheritance rather than on a single dominant mechanism.

\begin{landscape}
\begin{table*}
\flushleft
\caption{Comparative synthesis of conceptual frameworks for failure in geomaterials.}
\label{tab:frameworks}
\renewcommand{\arraystretch}{1.3}
\begin{tabularx}{21cm}{p{2.5cm} p{3.5cm} p{3.5cm} p{2cm} p{3.5cm} p{3.6cm}}
\hline
\textbf{Framework} & \textbf{Core mechanism} & \textbf{Typical \newline observables} & \textbf{Predictability} & \textbf{Strengths} & \textbf{Limitations} \\
\hline

Critical point / \newline damage mechanics 
& Progressive microcrack growth and coalescence in heterogeneous media leading to a critical state
& Accelerating displacement,  acoustic/seismic emissions, power-law scaling, log-periodicity 
& Moderate--high 
& Captures emergent system-scale behavior; explains precursors and scaling laws; unifies diverse signals 
& Non-unique signals; not all systems exhibit criticality; sensitive to noise \\

\hline
Rate-and-state friction / \newline frictional instability 
& Velocity-dependent friction and state evolution leading to slip instability 
& Slip rate acceleration, displacement time series, creep transitions 
& Moderate 
& Strong physical basis; explains stable creep $\rightarrow$ failure; mathematically well-developed 
& Assumes localized shear zone; \sout{distributed damage}; parameter calibration challenging \\

\hline

Threshold / \newline hydromechanical triggering 
& External forcing (e.g., rainfall, pore pressure) exceeds critical threshold reducing effective stress 
& Rainfall intensity-duration, pore pressure, groundwater levels 
& Low--moderate 
& Operational simplicity; widely used in early warning; strong empirical support 
& Neglects internal damage evolution; prone to false positives/negatives; site-specific thresholds \\

\hline

Structural / \newline kinematic controls 
& Failure governed by geometry, lithology, and pre-existing discontinuities 
& Slope geometry, fracture networks, structural mapping 
& Low (timing), high (susceptibility) 
& Identifies failure modes and locations; essential for hazard zoning 
& Static; does not capture temporal evolution or precursors \\

\hline

\end{tabularx}
\end{table*}
\end{landscape}

\section{A Unified Conceptual Framework}\label{framework}

\subsection{Architecture of the proposed framework}

\begin{figure}
\includegraphics[width=\textwidth]{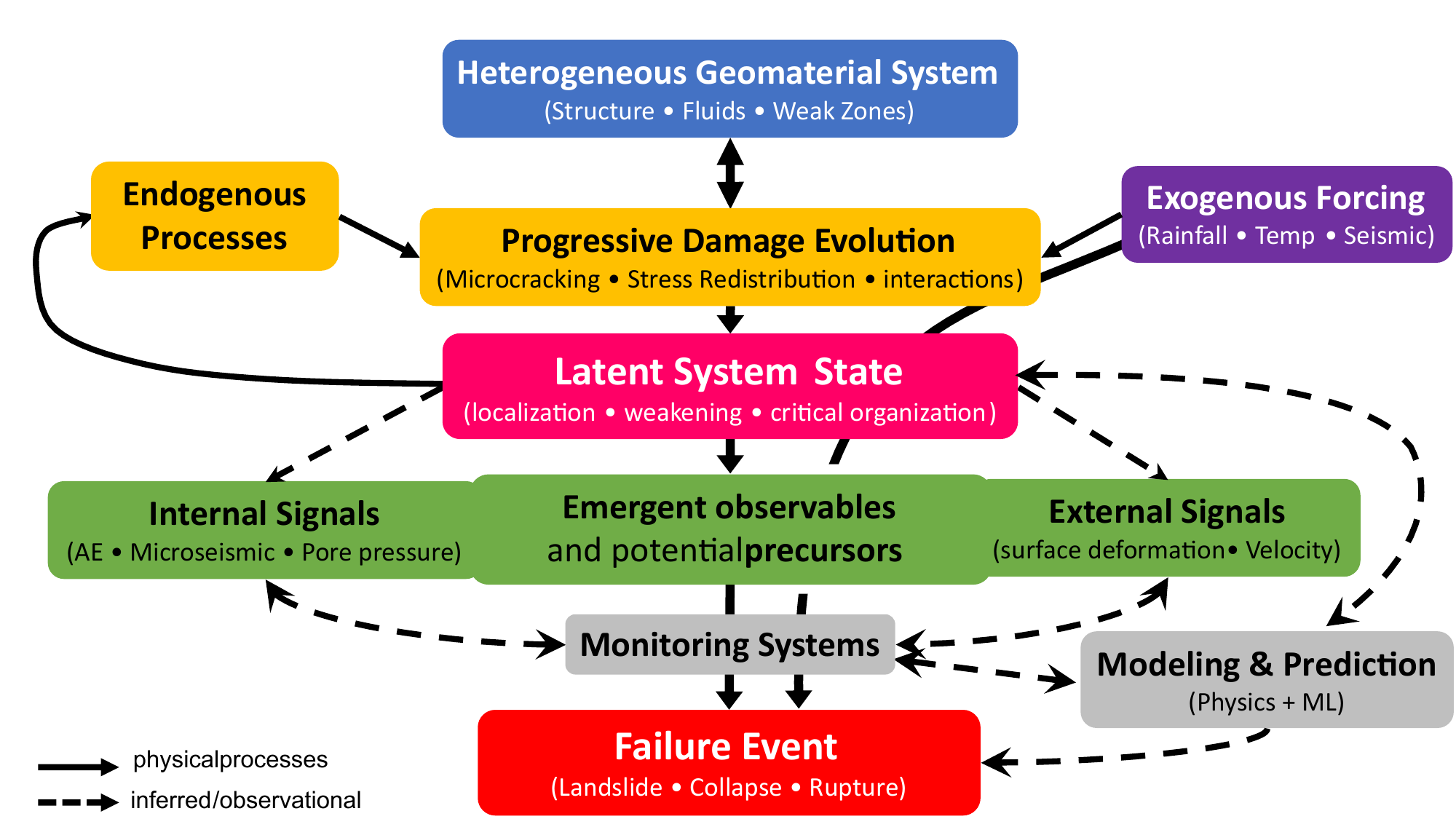}
\caption{Proposed unified conceptual framework for gravitational instabilities. Initial heterogeneity controls the development of progressive damage through endogenous processes that interact continuously with exogenous forcing. Damage evolution modifies the latent system state, which governs the emergence of observable signals sampled by monitoring systems and interpreted through modelling approaches. Failure occurs when progressive weakening and localization lead to loss of mechanical stability. The framework emphasizes that observations provide indirect information on system state rather than direct measurements of damage, making prediction fundamentally a problem of state estimation under uncertainty.}

\label{figframework}
\end{figure}

We propose a unified framework in which gravitational instability emerges from the progressive evolution of a heterogeneous geomaterial system toward failure (Figure~\ref{figframework}). The framework is organized around five interacting domains that collectively describe the transition from distributed damage to catastrophic rupture. Rather than viewing failure as a sudden event, the framework interprets it as the culmination of a continuous evolution linking material heterogeneity, damage accumulation, system-state evolution, observable signals, and eventual loss of mechanical stability.

The upper blue domain represents the heterogeneous geomaterial system, which defines the initial conditions governing subsequent evolution. This domain encompasses material heterogeneity, inherited damage, structural discontinuities, fluid distribution, and spatial variability in mechanical properties. These features control stress concentrations, influence stress redistribution, and determine where deformation and damage preferentially develop.

The central orange domain represents damage processes. Under gravitational loading, the system evolves through irreversible mechanisms such as microcrack nucleation and propagation, creep deformation, frictional weakening, grain debonding, and interface degradation. Through nonlinear feedbacks and stress redistribution, these processes progressively alter the mechanical integrity of the material and drive the system away from its initial state.

The cumulative effect of these interacting processes is represented by the central pink domain, corresponding to the evolving \emph{latent system state}. The latent system state is defined as the multidimensional internal mechanical condition of a heterogeneous geomaterial, integrating the cumulative effects of damage accumulation, stress redistribution, strain localization, frictional evolution, fluid interactions, and other processes controlling its stability. It provides a compact description of the evolving condition of the system and its proximity to instability. Although it cannot be observed directly, the latent system state governs the emergence of measurable signals and the overall trajectory toward failure.

As the system evolves, observable manifestations of its internal dynamics emerge. The green domain represents observable signals generated by the evolving system state, including deformation, displacement rates, microseismicity, acoustic emissions, hydrological variations, thermal anomalies, and other measurable quantities. These observables provide indirect and often non-unique information about the internal condition of the system, serving as proxies rather than direct measurements of damage processes.

Monitoring systems and modeling approaches, in grey in Figure~\ref{figframework}, aim to infer the evolving system state from observable signals and thereby constrain the system's proximity to instability.

The lower red domain represents macroscopic failure, which occurs when progressive weakening and localization lead to a loss of mechanical stability. At this stage, the system undergoes large-scale rupture, collapse, or gravitational failure. Failure is therefore interpreted as the final outcome of a continuous damage-to-failure trajectory governed by the interplay between heterogeneity, internal degradation processes, and external forcing.

\subsection{Endogenous evolution and exogenous forcing}

A central feature of the proposed framework is the coupling between endogenous processes and exogenous forcing, represented on both sides of Figure~\ref{figframework}. The temporal evolution of the system cannot generally be attributed to a single triggering mechanism, but rather emerges from interactions between internal degradation processes and external perturbations.

Endogenous processes correspond to the intrinsic evolution of the material under sustained or evolving loading conditions. These include damage accumulation, healing, stress redistribution, creep deformation, and progressive weakening. Such processes define the internal trajectory of the system toward instability.

Exogenous processes correspond to forcing mechanisms acting externally on the system, including hydrological forcing, temperature variations, seismic loading, and anthropogenic disturbances. These perturbations may modify either stress conditions or material properties, thereby accelerating or delaying destabilization. System response depends more on internal damage state than forcing amplitude alone.

\subsection{Emergence of observables and predictability}

An important implication of the framework is that the internal evolution toward failure may generate measurable observables represented in the lower part of Figure~\ref{figframework}. These observables do not directly measure damage itself, but instead provide indirect information on the evolving mechanical state of the system.

Within the framework, monitoring systems and modeling approaches act as tools for sampling and interpreting these evolving observables. 
The emergence of correlated and accelerating signals defines a potential window of predictability associated with the potential transition toward the critical state.

Importantly, this predictability remains inherently limited. Similar observable patterns may arise from distinct physical mechanisms or transient environmental perturbations, implying that observables are generally non-unique proxies of internal system state. Consequently, prediction must be interpreted probabilistically and within the broader physical context of system evolution.

\subsection{Scope and organization of the framework}

The proposed framework is intentionally generic and does not seek to replace system-specific models or hazard assessments. Rather, it provides a common conceptual structure for interpreting diverse gravitational instabilities through shared physical principles linking heterogeneity, damage evolution, system state, observables, and failure.

The following sections examine the different components of the framework successively: material heterogeneity (Section~\ref{Heterogen}), progressive damage evolution (Section~\ref{damage}), emergence of observable signals and precursors (Section~\ref{precursor}), monitoring strategies (Section~\ref{monitoring}), modeling approaches (Section~\ref{model}), and implications for prediction and early warning (Section~\ref{EWS}).

\section{Damage evolution and emergent precursors}

\subsection{Material Heterogeneity as necessary Condition}
\label{Heterogen}
Natural geomaterials such as rock, ice, and soil are intrinsically heterogeneous across a wide range of spatial and temporal scales. This heterogeneity originates from their formation processes and subsequent mechanical and environmental history \citep{Epes2022}, leading to grain-scale variability, the presence of fractures and discontinuities, and spatially variable water content. These features are not merely secondary complexities but fundamental properties that govern the mechanical behavior of geomaterials and their evolution toward failure  \citep{Guerriero2021}.

A key implication of this heterogeneity is that failure is typically controlled by the weakest elements within the material rather than by its average properties. From a theoretical point of view, this behavior arises because heterogeneous materials can be described as assemblies of elements with distributed strengths, where failure is governed by extreme-value statistics such that instability is triggered when the most critical element reaches its failure threshold \citep{Weibull1951,Zhang2013}. In the context of gravitational instabilities, this implies that rupture initiation is inherently localized within pre-existing or evolving weak zones, such as fractures, interfaces, or fluid-weakened regions \citep{Faillettaz&Or2015,Stead2015,Lan&al2022}.

At the grain scale, variations in mineral composition, crystal structure, and bonding conditions induce strong contrasts in local mechanical properties. At larger scales, discontinuities such as joints, bedding planes, or stratification introduce anisotropy and define preferential zones of weakness. In addition, the distribution of fluids—particularly water—adds a dynamic component by modifying effective stress, cohesion, and frictional resistance over time \citep{Guerriero2021}. 

Failure mechanisms in geomaterials span a continuum between brittle and ductile behavior. Brittle failure is characterized by rapid crack propagation with limited prior deformation, typically occurring in materials such as rock or cold ice under low-temperature or high strain-rate conditions \citep{Griffith1921, Anderson2017}. In contrast, ductile deformation involves significant strain accumulation prior to failure, allowing partial stress redistribution and delaying macroscopic rupture \citep{Arenson2007}. This behavior is commonly observed in soils or in ice under temperate conditions. However, this distinction remains conditional: the same material may exhibit either brittle or ductile responses depending on stress state, temperature, strain rate, and fluid conditions \citep{Mamot2018, Schulson2022}.

As a consequence, failure should be interpreted as the outcome of a progressive, multi-scale process rather than a purely instantaneous event. The coupling between heterogeneity, damage evolution, and stress redistribution governs both the timing and geometry of rupture. This inherent complexity introduces significant variability in failure behavior, which limits the applicability of purely deterministic approaches and motivates the development of probabilistic or physics-based descriptions of geomaterial failure.

Because stress redistribution occurs unevenly within heterogeneous materials, damage evolution becomes intrinsically localized, intermittent, and history-dependent. Understanding how these local degradation processes interact and progressively organize toward instability therefore requires examining the temporal dynamics of damage accumulation itself.

\subsection{Progressive endogenous damage and exogenous forcing}
\label{damage}

Failure in geomaterials results from the interaction between endogenous and exogenous processes \citep{Sornette2006}. Endogenous processes describe the internal evolution of the system through damage accumulation, stress redistribution, creep, and strain localization. Exogenous processes correspond to external perturbations such as hydrological, thermal, tectonic, or anthropogenic forcing. Within the proposed framework (Fig.~\ref{figframework}), instability emerges through the interaction of mechanisms operating across multiple spatial and temporal scales.

Endogenous evolution progressively modifies the mechanical state of the geomaterial. Damage develops through mechanisms such as microcracking, frictional sliding, creep deformation, grain debonding, and cohesion loss \citep{Griffith1921,Alava&al2006}. Although these processes operate at small scales, their cumulative effects alter the stress field and mechanical properties of the medium. Through stress redistribution, local damage promotes further weakening in neighboring regions, creating positive feedbacks that may lead to strain localization and the emergence of preferential rupture pathways \citep{Rudnicki1975,Ben-zion2008,Faillettaz&al2015a}. Failure therefore emerges not from an isolated defect but from the collective organization of interacting damage processes.

The rate and style of damage evolution depend strongly on material rheology and environmental conditions. Brittle systems may transition rapidly from localization to rupture, whereas ductile materials can accommodate prolonged deformation through progressive stress redistribution \citep{Amitrano2006}. Most natural systems occupy an intermediate regime controlled by factors such as strain rate, temperature, fluid pressure, confinement, and material composition.

Exogenous forcing influences this evolution by modifying boundary conditions, stress distributions, or material properties. Hydrological forcing increases pore pressure and reduces effective stress \citep{Iverson2000,Sidle2006}; thermal forcing promotes crack growth and material degradation through freeze--thaw and thermo-mechanical processes \citep{Kraublatter&al2013}; anthropogenic activities alter slope geometry and stress conditions \citep{Jaboyedoff2018}. These forcings may act progressively over long timescales or as transient perturbations over short timescales.

The response of a system to exogenous forcing depends strongly on its endogenous state. Identical rainfall events, thermal cycles, or seismic perturbations may produce markedly different outcomes depending on the degree of accumulated damage and antecedent weakening \citep{Fan&al2015,Faillettaz&al2010}. As endogenous degradation progresses, the system becomes increasingly sensitive to external disturbances, such that perturbations that are inconsequential in an intact material may trigger rapid destabilization in a highly damaged one \citep{Amitrano2006,Faillettaz&al2004}.

Instability therefore cannot be attributed solely to either endogenous damage or exogenous forcing. Failure emerges from their coupled evolution: endogenous processes progressively weaken the system, while exogenous forcing modulates and accelerates this evolution. Their interaction drives the transition from distributed deformation toward localized instability and ultimately rupture, while creating the conditions under which precursory signals may emerge \citep{Voight1988,Amitrano&al2005,Sornette2006}.

\begin{figure}
\includegraphics[width=\textwidth]{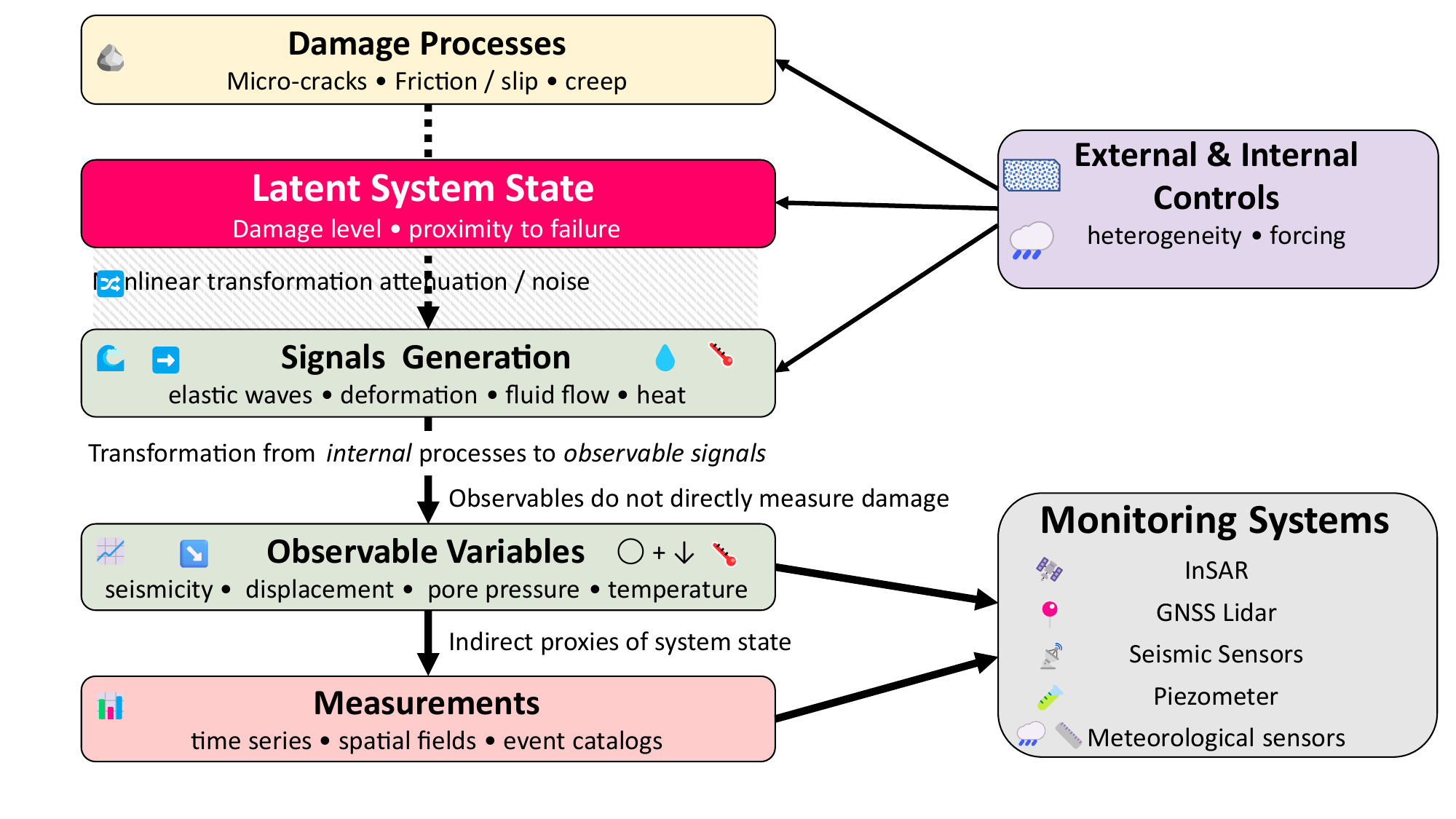}
\caption{Conceptual representation of the transformation from internal damage processes to observable signals and measurements: Progressive damage evolution modifies the latent system state, which reflects the internal mechanical condition of the geomaterial and its proximity to failure. Damage processes generate physical signals that are transformed into observable variables such as seismicity, deformation, pore pressure, and temperature, and subsequently recorded by monitoring systems. Because this transformation is influenced by heterogeneity, forcing conditions, attenuation, and noise, observations provide indirect and potentially non-unique information about the latent system state. Monitoring and interpretation therefore seek to reconstruct the evolving system state from incomplete observations rather than measure damage directly.}\label{figphysics}
\end{figure}

Although progressive damage governs the internal evolution of the system, these processes are rarely directly observable in natural environments. Instead, instability must be inferred through indirect measurements reflecting the evolving mechanical state of the medium. The transition from internal dynamics to measurable observables therefore represents a critical step in understanding predictability (Figure \ref{figphysics}). The nature and interpretation of these signals are examined in the following section.

\subsection{Emergent observables as proxies of system state}
\label{precursor}

The progressive evolution of damage is not directly observable but may be inferred through measurable signals that emerge from the evolving mechanical state of the system (Fig.~\ref{figframework}). These observables do not measure damage itself; rather, they reflect its consequences across different spatial and temporal scales (Fig. \ref{figphysics}). As damage accumulates, the statistical and dynamical properties of these signals may evolve, providing indirect information on system organization and proximity to instability.

Observables can be broadly distinguished according to whether they primarily sample internal damage processes or macroscopic system responses. Internal observables include acoustic emissions and microseismicity generated by microcrack nucleation, propagation, and coalescence \citep{Lockner1993}. Because they originate directly from damage processes, these signals are often sensitive to early-stage evolution. External observables correspond to the macroscopic expression of internal dynamics and include surface displacement, deformation rates, and morphological changes measured by geodetic or remote-sensing techniques. Hydrological and thermal variables occupy an intermediate position, as they both influence and reflect the evolving mechanical state through changes in pore pressure, fluid circulation, or temperature. Although these observables sample different scales and processes, they ultimately represent complementary manifestations of the same underlying damage evolution.

\subsection{Scaling, criticality and predictability}
As damage progresses, some observables may exhibit characteristic scaling behavior, particularly in systems approaching highly organized states prior to failure.
A common feature is the acceleration of activity prior to failure, observed in quantities such as deformation rates, microseismic event frequency, or energy release \citep{Faillettaz&al2019,Faillettaz&al2016}. At the event scale, rupture populations often display scale-invariant statistics, with the probability density of event sizes \(S\) following

\begin{equation}
\label{GR}
P(S) \sim S^{-\beta},
\end{equation}

where \(\beta\) characterizes the relative abundance of small and large events \citep{GutenbergRichter1944,Alava&al2006}. Such power-law distributions indicate the absence of a characteristic rupture scale and reflect the collective interaction of damage processes in heterogeneous media.

Temporal evolution may also exhibit scaling behavior. The acceleration of precursory activity is frequently described by

\begin{equation}
\label{eqPL}
X(t) \sim (t_c-t)^{-m},
\end{equation}

where \(X(t)\) is an observable, \(t_c\) the failure time, and \(m\) a scaling exponent \citep{Voight1988}. In some systems, this acceleration is further modulated by log-periodic oscillations,

\begin{equation}
\label{eqLogP}
X(t) \sim (t_c-t)^{-m}
\left[
1 + C\cos\left(\omega\ln(t_c-t)+\phi\right)
\right],
\end{equation}

where \(C\), \(\omega\), and \(\phi\) describe the amplitude, logarithmic frequency, and phase of the oscillations, respectively \citep{Faillettaz&al2008,Faillettaz&al2015a,Faillettaz&al2016b}. These patterns have been associated with hierarchical organization, cascading damage interactions, and discrete scale invariance \citep{Sornette1998,Sornette2006}, although their occurrence depends strongly on heterogeneity, forcing conditions, and observational scale. While their physical origin remains debated, they may provide additional structure and predictive information in the temporal evolution of precursory signals \citep{Lei&al2023,Lei2025}.

Such scaling laws are often interpreted as signatures of increasing interactions among damage processes and the emergence of collective behavior across scales.
Taken together, these scaling relationships suggest that observable signals are emergent expressions of collective damage evolution. They provide a bridge between internal system dynamics and measurable quantities, thereby forming the basis for monitoring, interpretation, and forecasting.

The emergence of accelerating activity, increasing correlations, and scale-invariant behavior is commonly interpreted within the framework of critical phenomena, in which failure corresponds to a transition from a relatively stable to an unstable system state \citep{Sornette2006}. In this perspective, progressive damage accumulation promotes increasing interactions among local failure processes, leading to the development of collective behavior across a broad range of spatial and temporal scales. Laboratory experiments and field observations have documented patterns consistent with such critical organization, including accelerating deformation, evolving event statistics, and growing spatial and temporal correlations \citep{Main1996,Sammis2002,Sornette2006}.

The development of collective behavior has important implications for predictability.  
As interactions among damage processes increase, observable signals may become increasingly sensitive to the evolving system state, potentially providing information on the system's proximity to instability and improve failure forecasting \citep{Voight1988,Amitrano&al2005,Faillettaz&al2016}.

Importantly, the framework proposed here does not assume that all gravitational instabilities evolve toward a critical state prior to failure. Criticality is regarded as one possible emergent regime within a broader spectrum of damage-to-failure trajectories. Similar signal patterns may arise from transient forcing, threshold effects, or non-stationary environmental conditions without necessarily reflecting the development of a critical transition.

Consequently, predictability remains fundamentally probabilistic. Observable signals provide indirect information on system evolution, but their interpretation is constrained by material heterogeneity, incomplete observations, and the non-uniqueness of signal--process relationships. Rather than enabling deterministic prediction, these observables primarily contribute to constraining the evolving system state and identifying changes in stability that may precede failure.

\subsection{Hidden criticality and multiple pathways to failure}
\label{sec}

Critical phenomena provide a powerful framework for understanding progressive failure in heterogeneous geomaterials. However, identifiable critical behavior is not systematically observed prior to catastrophic collapse. Many natural systems fail without clear precursory signatures, while others appear to be dominated by structural controls or external forcing. The apparent absence of precursors, however, should not be interpreted as evidence that progressive damage accumulation or collective organization are absent.

Several classes of instability illustrate this apparent discrepancy. In structurally controlled failures, such as planar rockslides, wedge failures, or fault reactivation, pre-existing discontinuities largely determine the geometry and location of rupture \citep{Stead2015}. Rapid brittle failures may occur over timescales too short for large-scale cooperative behavior to develop or to be detected, resulting in limited observable precursory activity \citep{Lockner1993}. Other instabilities are commonly described as threshold-driven, particularly rainfall-induced landslides, where transient hydrological forcing and pore-pressure increases exert a dominant control on failure timing \citep{Guzzetti2008,Gariano2016}. Similarly, externally triggered failures may result from earthquakes, volcanic activity, reservoir fluctuations, excavation, or blasting, for which the triggering perturbation provides the immediate impulse for collapse \citep{Keefer2002,Petley2012}.

These mechanisms are not necessarily alternative to progressive damage evolution. Structural controls, threshold exceedance, and external perturbations often determine the timing and mode of failure, but not necessarily the long-term evolution of the system toward instability. In many cases, failure may occur in a material that has already undergone substantial weakening through damage accumulation, stress redistribution, and progressive localization. The triggering process then acts on a system that is already close to rupture, even if its internal evolution remains unresolved by available observations.

This distinction highlights a fundamental difference between the existence of critical organization and its observability. The emergence of collective behavior depends on the internal dynamics of the system, whereas its observation depends on the monitored variables, sensor coverage, temporal resolution, and signal-to-noise ratio. Consequently, the apparent absence of critical precursors may reflect observational limitations as much as the underlying physics of failure.

Failure should therefore be viewed as the result of interactions among progressive damage, structural controls, and external forcing rather than as the outcome of mutually exclusive mechanisms. In some systems, damage evolution produces observable acceleration, increasing correlations, and enhanced predictability. In others, the same processes may remain hidden until collapse is initiated by threshold exceedance or external perturbation. Most natural hazards likely occupy intermediate positions along this continuum, exhibiting varying degrees of internal organization and forcing dependence.

From a monitoring perspective, the objective is therefore not solely to identify universal critical precursors, but to constrain the evolving state of the system and its proximity to instability. 
Within this framework, criticality is viewed not as a prerequisite for failure, but as one possible regime that can enhance predictability through the emergence of collective behavior. The primary objective of monitoring is therefore to constrain the evolving system state and its proximity to instability, regardless of whether critical dynamics are directly observable.

\section{Observing the System: Monitoring as sampling of the conceptual framework}
\label{monitoring}

Monitoring systems provide indirect observations of evolving system dynamics through deformation, seismicity, pore pressure, thermal anomalies, and surface changes. Combining these measurements helps constrain the transition from distributed damage to critical behavior and eventual failure.

Effective monitoring therefore requires combining observations that constrain the kinematic response of the system with those that provide insight into the underlying physical processes. External measurements primarily reflect the macroscopic evolution of deformation, whereas internal measurements provide proxies for damage evolution and state variables controlling stability. Their joint interpretation allows observations to be mapped onto different stages of system evolution, from distributed damage to critical state and eventual failure.

\subsection{External parameters: surface monitoring}

Surface monitoring techniques primarily sample the macroscopic response of the system, corresponding to the external expression of internal processes within the conceptual framework. These observations provide spatially distributed information on deformation patterns, enabling the identification of active zones and the characterization of system kinematics \citep[for a review]{Thirugnanam&al2022}.

Among these, Interferometric Synthetic Aperture Radar (InSAR) has become a key tool for detecting and mapping ground deformation over large areas with high spatial coverage \citep{Meng&al2024}. Its ability to provide displacement time series makes it particularly suitable for tracking the temporal evolution of slow-moving instabilities and identifying phases of acceleration associated with transitions toward critical behavior. 
However, InSAR observations are subject to important limitations. Measurements correspond only to displacement projected along the satellite line of sight rather than the full three-dimensional displacement field, and are derived from spatially coherent areas, effectively averaging deformation over resolution cells. Consequently, InSAR primarily captures broad-scale surface deformation and may underestimate localized or heterogeneous motions. In addition, the method requires sufficient phase coherence between successive acquisitions, so dense vegetation, rapid surface changes, snow cover, or strong decorrelation can substantially reduce measurement quality or prevent reliable displacement retrieval.

Complementary approaches such as LiDAR enable high-resolution topographic surveys, allowing precise quantification of surface changes and volumetric estimates of mass movement \citep{Jaboyedoff&al2012}. Ground-based radar systems further extend these capabilities by providing high-frequency measurements of displacement, which are especially valuable for capturing rapid accelerations and short-term variations \citep{Casagli&al2023}.

These techniques primarily constrain the observable consequences of damage evolution, rather than the damage processes themselves. As such, they are particularly effective for identifying late-stage evolution and the approach to failure, but provide limited direct insight into subsurface mechanisms.

\subsection{Internal parameter monitoring}

Internal monitoring techniques sample processes occurring within the material and therefore provide some of the most direct constraints on the evolving mechanical state of a system. These observations are particularly valuable because they provide information on damage accumulation, stress redistribution, and changes in material properties that are not always detectable from external parameter monitoring alone.

Seismic measurements constitute a major source of information on internal system evolution. Event-based observations, including acoustic emissions and microseismicity, record the release of elastic energy associated with microcrack nucleation, propagation, and coalescence \citep{Amitrano&al2005}. Changes in event rates, magnitudes, and spatial clustering may therefore provide insight into the organization and localization of damage during the approach to failure.

Complementary information can be obtained from wavefield-based approaches such as ambient-noise seismic interferometry. Temporal variations in seismic wave velocity (\(dv/v\)) reflect changes in the elastic properties of the medium associated with stress redistribution, crack opening and closure, fluid circulation, and progressive damage accumulation \citep{Brenguier&al2008,Larose&al2015,LeBreton&al2021}. Unlike event-based monitoring, which records discrete rupture processes, \(dv/v\) measurements provide a continuous estimate of the evolving state of the material. Recent studies have demonstrated their ability to detect seasonal changes in slope stability and subtle precursory modifications of rock masses prior to failure \citep{Weber&al2018}, highlighting their potential for tracking progressive weakening in heterogeneous geomaterials.

Hydrological measurements provide another important window into internal system dynamics. Pore pressure directly controls effective stress and therefore strongly influences the balance between resisting and driving forces \citep{Iverson2000}. Temporal variations in pore pressure can both reflect and accelerate ongoing damage processes, particularly in systems where fluid circulation plays a dominant role in destabilization. Similarly, temperature measurements are increasingly used in environments where thermo-mechanical coupling is important, such as permafrost and glacierized slopes. Through their influence on material properties, ice content, and phase transitions, temperature variations can significantly affect damage evolution and system sensitivity to external forcing \citep{Li&Yin2025}.

Despite their value, internal observations remain inherently limited by sensor coverage and accessibility. They often provide detailed information at specific locations while only partially sampling the spatial heterogeneity of the system. Consequently, their interpretation requires careful consideration of representativeness, scale effects, and the relationship between local measurements and system-scale behavior.

\subsection{Multi-sensor integration}

Because individual monitoring techniques only provide partial views of the system, the integration of multiple datasets is essential to adequately sample the different components of the conceptual framework \citep{Whiteley&al2019,Casagli&al2023,Uhlemann&al2016}. By combining internal and external observations, it becomes possible to link damage evolution, state variables, and macroscopic response within a unified interpretation.

\begin{figure}
\includegraphics[width=\textwidth]{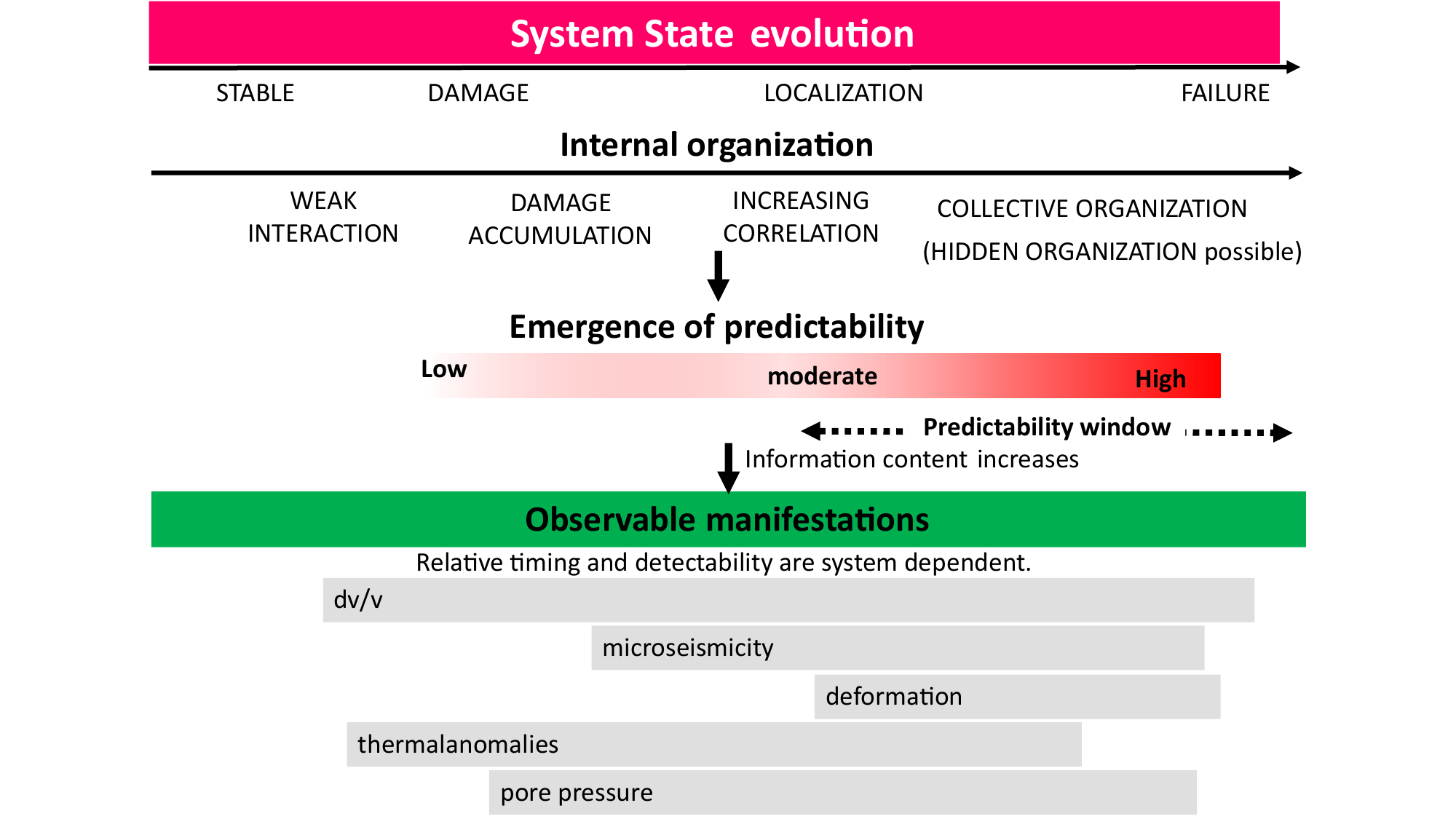}
\caption{Conceptual representation of the emergence of predictability during the evolution of gravitational instabilities. Progressive damage accumulation and increasing interactions may generate enhanced correlations and, in some systems, collective organization. This increasing organization may produce a potential window of predictability through the emergence of observable manifestations that provide indirect and imperfect information on the evolving system state. The examples shown are illustrative rather than exhaustive, and future observations or indicators may reveal additional manifestations of system evolution. Because observability depends on the monitored variables, sensor coverage, resolution, and signal-processing methods, collective organization may remain hidden and some systems may fail without detectable precursory signatures. Thus, predictability is interpreted as an emergent and probabilistic property rather than a universal feature of approaching failure.}\label{figprec}
\end{figure}

This multi-sensor approach improves the detection of early signals by identifying consistent patterns across independent measurements and by capturing transitions between stages of system evolution \citep{Intrieri2019}. For example, the co-occurrence of increasing microseismic activity, pore pressure variations, and accelerating surface displacement may indicate the transition from distributed damage to a localized rupture pathway leading to catastrophic instability \citep{Faillettaz&al2015a}(Fig. \ref{figprec}).

In addition, redundancy across datasets enhances the robustness of interpretation and reduces the risk of misattributing signals to incorrect processes. However, multi-sensor integration remains challenging due to differences in spatial and temporal resolution, data quality, and measurement sensitivity \citep{Samadzadegan&al2025}. Furthermore, the relationships between observed signals and underlying processes are often non-unique, complicating direct inference of system state.

Addressing these challenges requires the development of integrated approaches that explicitly link observations to the conceptual framework, often through physics-based or probabilistic models \citep{Intrieri2019,Peng2014}. Such approaches aim to translate heterogeneous datasets into coherent estimates of system evolution, ultimately improving both understanding and prediction of gravitational instabilities.

\section{Modeling approaches across the damage-to-failure spectrum}
\label{model}
Modeling approaches can be interpreted according to the specific components of the system they represent, from progressive damage evolution to the emergence of criticality and eventual failure. The principal modeling approaches and the stages of the framework they represent are summarized in Table~\ref{tab:models}, while their conceptual relationships are illustrated in Figure~\ref{figuremodel}. Rather than providing a single unified description, different model classes capture complementary aspects of this evolution, reflecting trade-offs between physical realism and conceptual abstraction.

\subsection{Continuum approaches: deformation and stress evolution}

Continuum approaches primarily represent the mechanical evolution of the system prior to failure, including stress redistribution, deformation, and strain localization. In these models, geomaterials are treated as continuous media governed by constitutive laws, and the governing equations of continuum mechanics are solved using numerical methods such as the finite element method (FEM), finite difference method, or material point method \citep{Jing2003,Stead&al2006}.

These approaches are particularly well suited to capturing the progressive damage stage of the framework, where deformation remains spatially distributed and controlled by material properties and boundary conditions . When coupled with advanced constitutive laws (e.g., elastoplasticity, viscoplasticity, or damage mechanics), they allow detailed simulation of stress–strain evolution and the onset of localization \citep{Soga&al2015}.

However, their ability to represent the transition toward failure is limited by the assumption of continuity. Discontinuities such as cracks, joints, or fragmentation must be introduced through additional formulations, which may not fully capture the discrete nature of rupture processes. This limitation becomes critical near failure, where the system transitions from distributed deformation to localized rupture and block interaction \citep{Wei&al2025}.

Despite these challenges, continuum models remain essential for site-specific applications due to their strong physical basis and their ability to integrate multiphysics processes, particularly in the early to intermediate stages of system evolution.

\subsection{Discrete and hybrid approaches: localization and rupture dynamics}

Discrete and hybrid approaches extend the framework by explicitly representing discontinuities and interactions between discrete elements. In discrete element models (DEM), the material is represented as an assembly of interacting particles or blocks, allowing direct simulation of crack propagation, fragmentation, and post-failure dynamics.

These approaches are particularly well suited to capturing the transition from localized damage to macroscopic failure, where block interactions and large deformations dominate system behavior. Hybrid approaches (e.g., FEM--DEM) combine continuum descriptions of intact material with discrete representations of fractured zones, enabling a more complete description of the system across multiple stages of the framework \citep{Jing2003,Stead&al2006,Desrues&al2019, Liang2019, Dai&al2025}.

While these models provide a more realistic representation of rupture processes, they are computationally demanding and require extensive parameterization, which may limit their applicability at large scales.

\subsection{Statistical physics models: emergence of criticality and precursors}

\begin{figure}
\includegraphics[width=\textwidth]{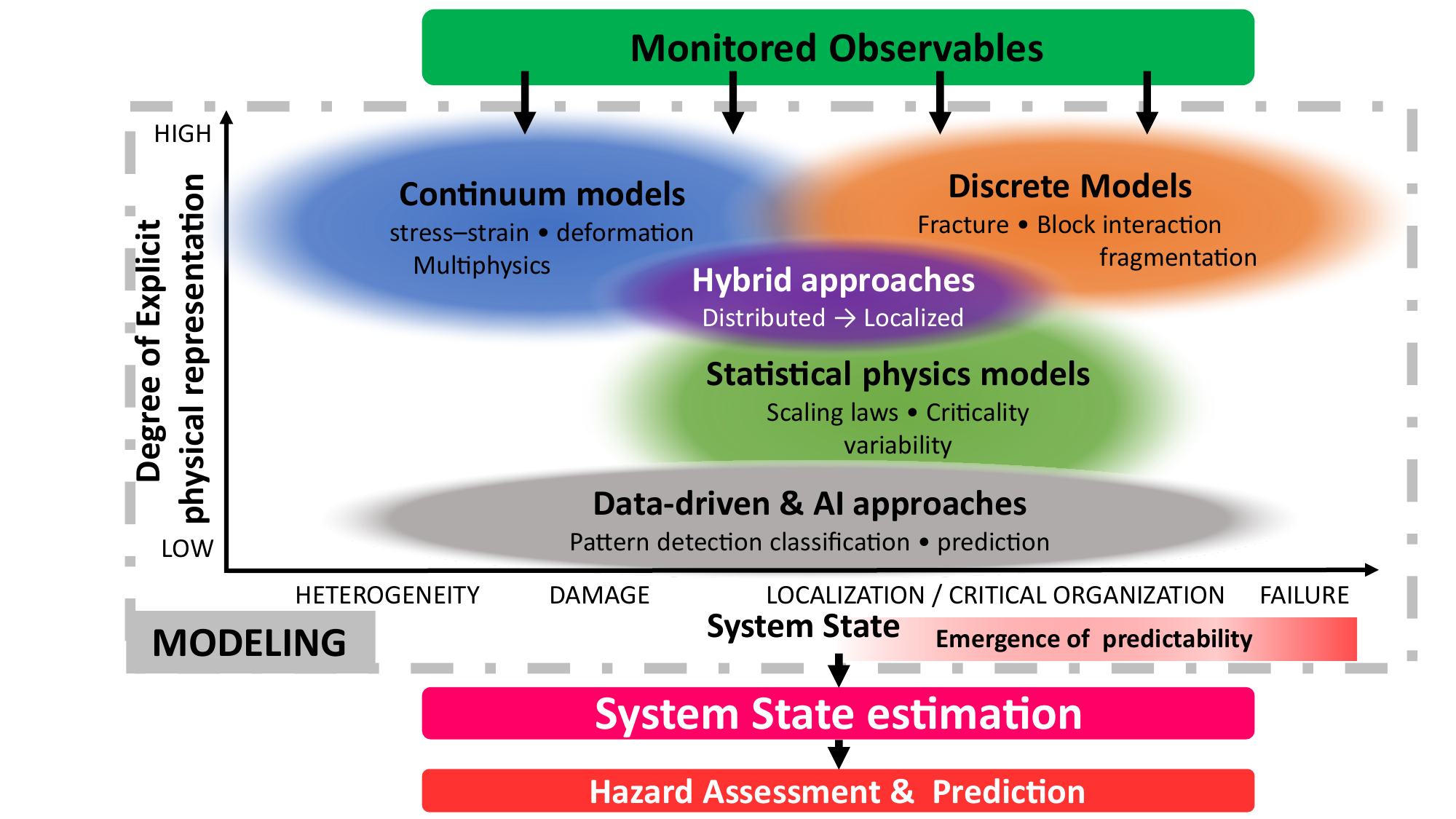}
\caption{Schematic representation of the complementarity of modeling approaches across the damage-to-failure spectrum. The horizontal axis represents system evolution from heterogeneous material to failure through the evolving system state, whereas the vertical axis reflects the degree of explicit physical representation. Continuum models primarily describe distributed deformation and multiphysics processes, discrete and hybrid approaches capture localization and rupture, statistical physics models explain scaling and collective organization, and data-driven approaches map monitored observables onto latent system states. Together, these approaches contribute to system-state estimation and forecasting, while predictability generally emerges as instability develops.}\label{figuremodel}
\end{figure}

Statistical physics models provide a complementary perspective by focusing on the collective behavior of heterogeneous systems approaching failure. These models are particularly suited to representing the emergence of criticality, where interactions between local damage events lead to correlated, system-wide behavior (Fig \ref{figuremodel}).

In statistical models of fracture such as fiber bundle models (FBM) or spring-block models (SBM), the material is represented as an ensemble of interacting elements with distributed properties \citep{Alava&al2006,Kawamura&al2012}. These models reproduce key features of the framework, including progressive damage accumulation, accelerating activity, and critical transitions \citep{Kun2006,Pradhan&al2010}.

A major strength of these approaches lies in their ability to explain the emergence of scaling laws, power-law statistics, and log-periodic patterns observed in precursory signals. They provide a natural framework for interpreting the transition from distributed damage to critical state, where correlations increase and system behavior becomes increasingly coherent.

However, their simplified representation of geometry and boundary conditions limits their direct applicability to site-specific prediction \citep{Faillettaz&al2012}. Instead, they are best viewed as conceptual and statistical tools for identifying generic behaviors and guiding the interpretation of observed signals.

\subsection{Data-driven and hybrid frameworks: mapping observables to system state}

Recent developments increasingly aim to bridge the gap between observations and physical models through data-driven and hybrid approaches \citep{Reichstein2019}. These methods focus on mapping observable signals onto the underlying system state within the conceptual framework, often without explicitly resolving all physical processes \citep{Lin2025,Jiang&al2026}.

Machine learning and statistical inference techniques are particularly effective at identifying patterns in large datasets, including precursory signals associated with the approach to failure. When combined with physics-based models, they offer a promising pathway to integrate multiple sources of information and improve the detection of critical transitions \citep{Yuan&al2025,Zhao&al2024}.

In this context, statistical physics models often serve as "synthetic laboratories" for generating training data, while continuum and discrete models provide physically grounded constraints. These hybrid approaches therefore act as integrative tools, linking observations, processes, and predictions within a unified framework.

\begin{landscape}
\begin{table*}[t]
\centering

\caption{Comparison of major modeling approaches according to the stages of the damage-to-failure framework they represent. Rather than providing a unified description of instability processes, different model classes capture complementary aspects of system evolution, from distributed deformation and damage accumulation to rupture localization, critical transitions, and observable precursor dynamics.}

\label{tab:models}

\renewcommand{\arraystretch}{1.3}

\begin{tabularx}{\linewidth}{
p{3cm}
p{4cm}
p{3.8cm}
p{4.5cm}
p{4.2cm}
}

\toprule

\textbf{Modeling approach} &
\textbf{Representation of \newline the system} &
\textbf{Stage of framework \newline primarily captured} &
\textbf{Key capabilities} &
\textbf{Main limitations} \\

\midrule

Continuum \newline approaches
&
Continuous medium governed by constitutive laws and continuum mechanics
&
Distributed deformation, stress redistribution, strain localization
&
Simulate stress--strain evolution; integrate multiphysics coupling; suitable for site-scale analyses
&
Limited representation of discontinuities, fragmentation, and rupture dynamics near failure \\

\midrule

Discrete and hybrid approaches
&
Assemblies of interacting blocks or particles, sometimes coupled with continuum domains
&
Localization of damage, crack propagation, rupture, post-failure dynamics
&
Capture fragmentation, block interaction, large deformation, and evolving discontinuities
&
High computational cost; extensive parameterization; scale limitations \\

\midrule

Statistical physics models
&
Ensembles of interacting heterogeneous elements with simplified geometries
&
Emergence of criticality, accelerating activity, correlated damage evolution
&
Reproduce scaling laws, power-law acceleration, critical transitions, and precursor statistics
&
Simplified geometry and boundary conditions; limited site-specific applicability \\

\midrule

Data-driven and \newline hybrid frameworks
&
Statistical inference and machine learning applied to observational and synthetic datasets
&
Mapping observables to system state and identifying precursory patterns
&
Integrate multi-sensor observations; detect hidden patterns; combine physical and statistical information
&
Limited interpretability; dependent on training data quality and representativeness \\

\bottomrule

\end{tabularx}

\end{table*}

\end{landscape}

\section{Prediction and early warning under uncertainty}
\label{EWS}

\begin{figure}
\includegraphics[width=\textwidth]{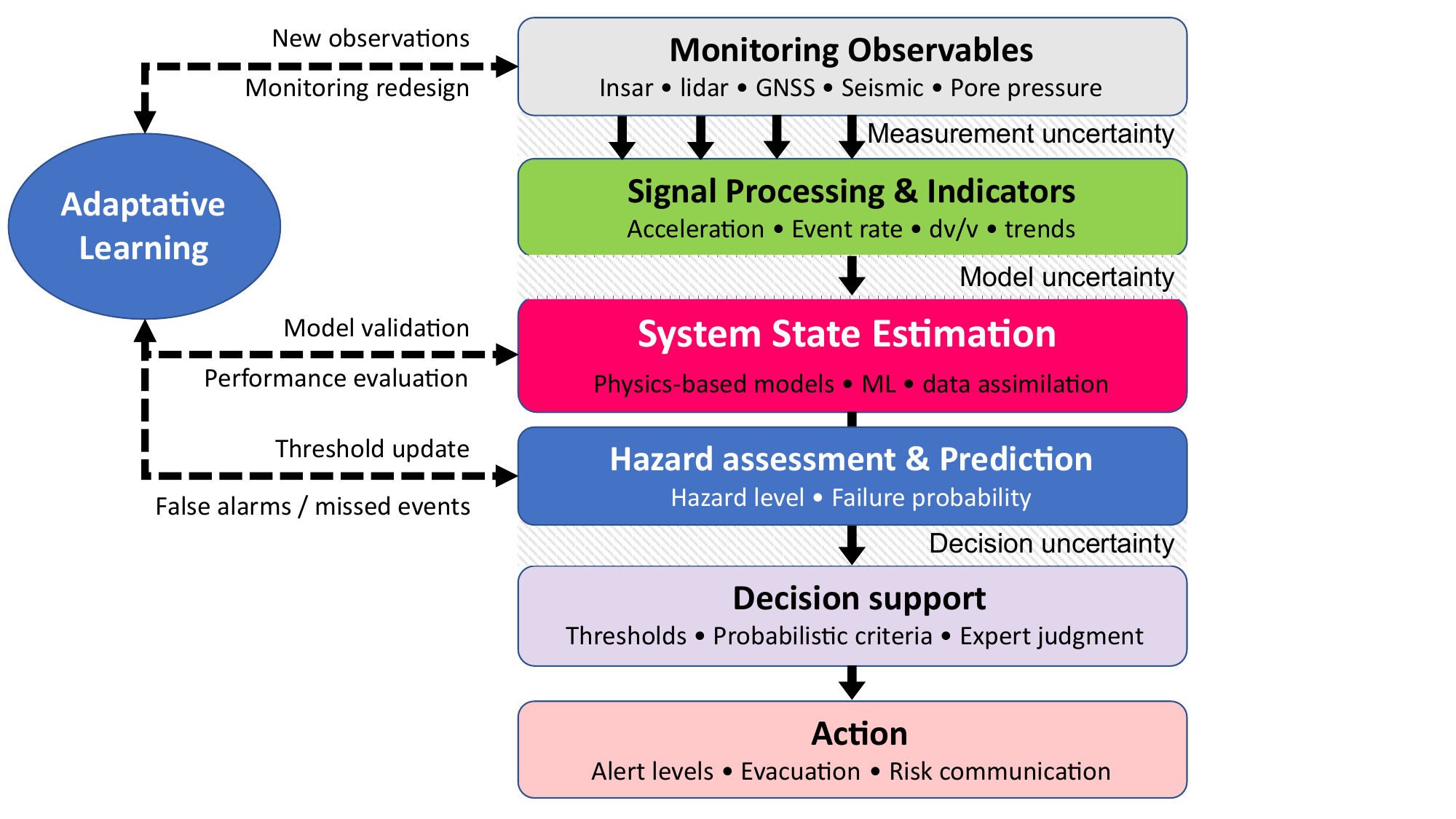}
\caption{Conceptual architecture of prediction and early warning for gravitational instabilities. Monitoring systems provide heterogeneous observations from which indicators are extracted and integrated to estimate the evolving system state. Physics-based, data-driven, and data-assimilation approaches are used to infer hazard levels and failure probabilities under uncertainty. These predictions support operational decision-making through thresholds, probabilistic criteria, and expert judgment, ultimately leading to actions such as alerts, evacuations, or risk communication. New observations and system outcomes continuously contribute to adaptive learning through model validation, performance evaluation, threshold revision, and optimization of monitoring strategies. Measurement, model, and decision uncertainties affect each stage of the process.}\label{figEWSconcept}
\end{figure}

Early warning systems (EWS) aim to detect and interpret the transition from progressive damage to a catastrophic failure (as expressed in Fig. \ref{figframework}). This transition defines the onset of practical predictability, where system dynamics become increasingly organized and potentially observable through evolving signals (Section \ref{damage}) and monitoring data (Section \ref{monitoring}). EWS therefore rely on the integration of observations, models, and interpretation strategies to translate partial measurements into actionable information (Fig \ref{figEWSconcept}).

\subsection{From observation to decision}

A central challenge for early warning systems is translating observations into operational decisions. Monitoring systems provide heterogeneous data streams that sample different components of system evolution, but these observations must be interpreted within a coherent framework to assess the level of instability.

Early warning systems transform heterogeneous observations into indicators of evolving instability. These indicators are then used to define thresholds or probabilistic criteria for issuing warnings. Such indicators can be interpreted as proxies for transitions between stages, particularly the shift from distributed damage to critical behavior.

Decision-making further involves translating these indicators into operational actions, often under constraints of uncertainty, limited data, and varying temporal scales. This process typically relies on expert judgment, empirical thresholds, or model-based forecasts, highlighting the need for transparent and physically grounded interpretation schemes.

\subsection{Challenges: uncertainty, non-uniqueness, and false alarms}

Despite significant progress, the development of reliable early warning systems remains constrained by several fundamental challenges. A primary difficulty arises from the indirect nature of observations: measurable signals do not uniquely correspond to specific internal processes, and similar patterns may emerge from different mechanisms or transient perturbations \citep{Sornette2002}.

Non-uniqueness complicates precursor interpretation and increases the risk of false alarms and missed events \citep{Staehli&al2015,Piciullo2020}. In particular, acceleration patterns or increases in activity may reflect external forcing rather than an intrinsic approach to failure \citep{Crosta2017}. Conversely, some systems may fail without clear or detectable precursors, especially when monitoring resolution is limited.

Uncertainty is further compounded by system heterogeneity, incomplete data coverage, and scale-dependent effects (Figure \ref{figEWSconcept}). These factors limit the ability to define universal thresholds or deterministic prediction rules \citep{Piciullo2020}. Early warning systems should therefore be viewed as probabilistic tools that assess evolving hazard levels rather than predict exact failure times.

\subsection{Emerging approaches: integration, seismic co-detection, extreme-event and nowcasting frameworks}

Recent developments aim to address these challenges by combining multiple sources of information and leveraging advances in data analysis. Multi-sensor integration and co-detection strategies seek to identify consistent patterns across independent datasets, thereby increasing the robustness of precursor identification and reducing ambiguity \citep{Faillettaz&al2016,Faillettaz&al2019}. Such seismic co-detection approaches may also prove particularly valuable for brittle rupture processes, where networks of sensors can enhance the detection of clustered and propagating rupture activity, even when individual precursory signals remain weak or spatially localized \citep{Faillettaz2023}.

In parallel, machine learning and data-driven approaches are increasingly used to detect complex patterns in large datasets and to map observations onto system states \citep{Reichstein2019}. When combined with physics-based models, these methods offer a promising avenue for integrating heterogeneous data and improving the detection of critical transitions.

Conceptual advances from statistical physics also provide new perspectives. The identification of ''dragon-kings'' (extreme events that deviate from standard scaling behavior) has been proposed as a potential indicator of system-wide transitions \citep{Sornette2006,Lei&al2023,Lei2025}. Similarly, record-breaking statistics and related approaches offer robust tools for detecting accelerating activity and approaching instability in noisy time series \citep{Kadar&al2020,Kadar&al2022}.

Another promising development is the application of nowcasting approaches, originally introduced in statistical seismology \citep{Rundle&al2021}, which shift the emphasis from predicting the exact timing of failure toward continuous assessment of a system's evolving condition. Rather than forecasting the next large rupture directly, nowcasting provides a probabilistic estimate of the system's progression toward instability based on continuously updated observations. Within the conceptual framework proposed here, this perspective is particularly compelling because it is conceptually consistent with viewing prediction as a problem of latent system-state estimation under uncertainty. Similar ideas also underpin recent landslide nowcasting systems that integrate heterogeneous environmental observations through data-driven approaches to assess evolving hazard conditions \citep{Stanley2021}.

Collectively, these developments illustrate a broader paradigm shift from the search for universal deterministic precursors toward probabilistic reconstruction of the evolving condition of gravitational systems through the integration of physical understanding, continuous observations, and data-driven inference.

\subsection{Limits and Remaining challenges}
Despite its integrative perspective, the proposed framework remains subject to important limitations that constrain both its generality and predictive applicability.

A first limitation arises from the intrinsic heterogeneity and multi-scale nature of geomaterials. 
Gravitational instabilities result from nonlinear interactions among mechanical, hydrological, thermal, and structural processes whose relative importance varies strongly between systems and through time \citep{Iverson2000,Hungr2014}. The framework should therefore be viewed as a generic interpretative structure rather than a universal deterministic model.

A second limitation concerns observability. Most monitoring techniques only provide indirect proxies of the internal mechanical state through deformation, seismicity, pore pressure, or thermal anomalies. Because similar observable patterns may result from distinct physical mechanisms, precursor signals remain fundamentally non-unique and often ambiguous in their physical interpretation \citep{Amitrano&al2005}. Moreover, some systems may fail with weak and/or unobservable precursory activity, limiting deterministic predictability.

Bridging scales also remains challenging. Laboratory experiments, simplified models, and numerical simulations reproduce important features of damage evolution and criticality, but their extrapolation to natural systems is limited by complex geometries, evolving boundary conditions, and incomplete knowledge of subsurface structure \citep{Main2017,Sornette2006}. Consequently, catastrophic failure does not always evolve through ideal critical transitions. Scaling laws and precursor dynamics identified in idealized systems may therefore not systematically translate to natural field conditions.

Beyond scientific and technical limitations, operational prediction also involves important societal and decision-making constraints. In practice, the objective of early warning systems is not only to detect instability, but also to support decisions such as evacuation or infrastructure closure under conditions of uncertainty. Determining the appropriate timing of intervention remains particularly challenging: warnings issued too late may fail to prevent casualties, whereas warnings issued too early or too frequently may generate unnecessary economic disruption, loss of credibility, and reduced social acceptability of future alerts. This tension reflects a fundamental trade-off between risk reduction and societal and institutional tolerance to false alarms  \citep{Eiser&al2012}. Consequently, the effectiveness of early warning systems depends not only on predictive performance, but also on the integration of uncertainty communication, decision strategies, and human factors within operational frameworks.

\section{Conclusions and Perspectives}
\label{conclusion}

Gravitational instabilities arise from the progressive evolution of heterogeneous geomaterials through multiple interacting physical processes operating across a wide range of spatial and temporal scales. This review has proposed a unified conceptual framework in which these processes are interpreted through the evolution of a latent system state, representing the internal mechanical condition of the system. Within this framework, catastrophic failure emerges as the culmination of a continuous evolution driven by damage accumulation, stress redistribution, localization, and external forcing, rather than as an isolated event.

The principal contribution of this framework is to provide a common language linking physical processes, observations, monitoring strategies, modeling approaches, and forecasting. Observable quantities do not directly measure damage or stability; instead, they provide indirect and incomplete information about the evolving latent system state. Monitoring systems, physical models, statistical approaches, and data-driven methods can therefore be viewed as complementary tools for estimating this hidden state and its proximity to instability. From this perspective, the central challenge of prediction shifts from identifying universal precursors to reconstructing the evolving system state from heterogeneous observations under uncertainty.

Despite this progress, significant challenges remain. The intrinsic heterogeneity of geomaterials, the indirect nature of observations, and the non-uniqueness of signal–process relationships limit predictability and complicate interpretation. These limitations highlight the need to move beyond deterministic approaches toward probabilistic and integrative frameworks that explicitly account for uncertainty.

Future research should focus on three key directions. First, advancing multi-scale coupling between models and observations is essential to bridge the gap between local damage processes and system-scale behavior. This includes linking micro-mechanical processes, meso-scale interactions, and large-scale deformation within consistent frameworks. Second, the development of real-time monitoring and analysis systems represents a critical step toward operational prediction. Integrating continuous data streams, automated processing, and adaptive models will enable more responsive and robust early warning strategies. Third, improving uncertainty quantification is fundamental to enhancing the reliability of forecasts. This requires better characterization of parameter variability, model uncertainty, and observational limitations, as well as the development of probabilistic approaches that explicitly incorporate these sources of uncertainty.

More broadly, the similarities observed across different types of gravitational instabilities suggest that they may belong to a wider class of failure phenomena in complex systems. Progressive damage accumulation, the emergence of correlations, scaling behavior, and, in some cases, critical-like transitions appear to be recurrent features across a range of natural and engineered systems. However, the proposed framework does not assume that all gravitational instabilities evolve through a critical state prior to failure. Rather, criticality is interpreted as one possible emergent regime within a broader spectrum of damage-to-failure trajectories. This perspective opens the possibility of developing a more general theory of failure grounded in common physical principles while preserving the diversity of mechanisms observed in natural systems.

Improving prediction will require moving beyond isolated disciplinary approaches toward integrated descriptions of evolving natural systems. Future progress will depend on the ability to couple high-resolution observations, multi-scale physics, statistical descriptions of failure, and adaptive data-driven inference within coherent predictive frameworks. In this perspective, gravitational hazards become not only geotechnical problems, but fundamental examples of critical transitions in complex heterogeneous systems. Developing such a generalized theory of failure represents both a major scientific challenge and a key societal objective in a rapidly changing environment.

\end{document}